\documentclass{aa}
\usepackage{graphicx}
\usepackage{psfig}
\usepackage{txfonts}
\def\ltsim{\raise 2pt \hbox {$<$} \kern-1.1em \lower 4pt \hbox {$\sim$}}
\def\ltapprox{\raise 2pt \hbox {$<$} \kern-1.1em \lower 5pt \hbox {$\approx$}}
\def\gtsim{\raise 2pt \hbox {$>$} \kern-1.1em \lower 4pt \hbox {$\sim$}}
\def\gtapprox{\raise 2pt \hbox {$>$} \kern-1.1em \lower 5pt \hbox {$\approx$}}

\def\arcsec{$^{\prime\prime}$}
\def\arcmin{$^{\prime}$}
\def\degrees{$^{\circ}$}

\begin{document}
%
%
   \title{XMM-Newton observations of the Coma cluster relic 1253+275 }

   \author{Luigina Feretti
          \inst{1,2}
          \and
          Doris M. Neumann\inst{2}
          }

   \offprints{L. Feretti}

   \institute{Istituto di Radioastronomia -- INAF, 
              via P. Gobetti 101, I--40129, Bologna, Italy\\
              \email{lferetti@ira.inaf.it}
         \and
AIM - UMR N$^\circ$ 7158 CEA - CNRS - Universit\'e Paris VII, 
DSM/DAPNIA/Service d'Astrophysique, CEA/Saclay, 
L'Orme des Merisiers, B\^at. 709, 91191 Gif-sur-Yvette, France
             \email{dneumann@cea.fr}
}


  \abstract 
{} 
{Using XMM Newton data, we investigate the nature of the X-ray emission in 
the radio relic 1253+275 in the Coma cluster.
 We determine the conditions of the  cluster gas to check current models 
of relic formation, and we  
set constraints  on the intracluster magnetic field.
  } 
  {Both imaging and spectral analysis are performed, and the X-ray emission
  is compared with the radio emission.  }  
{We found that the emission is of thermal origin and
  is connected to the sub-group around NGC~4839.  The best-fit gas
  temperature in the region of the relic and in its vicinity is in the
  range 2.8 $-$ 4.0 keV, comparable to the temperature of the NGC~4839
  sub-group.  We do not detect any high temperature gas, resulting from  a
  possible shock in the region of the Coma relic.  We therefore suggest
  that the main source of energy for particles radiating in the radio
  relic is likely to be turbulence.
From the X-ray data, we can also set a flux upper
limit  of $  3.2\times 10^{-13}$ erg cm$^{-2}$ s$^{-1}$,
in the 0.3 $-$ 10 keV energy range, to the
non-thermal emission in the relic region.  This 
leads to a 
magnetic field B $ > $ 1.05 $\mu$G.
}
{}

  \maketitle

\section{Introduction}

Some clusters of galaxies show diffuse extended radio emission in
their outskirts, indicating the presence of relativistic particles and
magnetic fields in the intracluster medium (ICM). These radio sources
are commonly called radio relics. Their steep radio spectra indicate
that the radiating particles have short lifetimes ($\sim$ 10$^8$ yr),
linked to synchrotron and inverse Compton (IC) radiation losses, and
thus are reaccelerated by some mechanism that acts with an
efficiency comparable to the energy loss processes.

Current theoretical models predict that relativistic particles
radiating in radio relics are powered by energy dissipated in shock
waves produced during the cluster formation. This picture is supported
by recent numerical simulations on cluster mergers (Ryu et al. 2003),
in which shock waves are present at the cluster periphery.  It has
been suggested that the electron acceleration required to produce the
relic emission  results from Fermi-I diffusive
shock acceleration of thermal ICM electrons (En{\ss}lin et al. 1998)
or by the adiabatic energization of relativistic electrons confined in
fossil radio plasma that was released by a formerly active radio galaxy
(En{\ss}lin \& Gopal-Krishna 2001, En{\ss}lin \& Br\"uggen 2002).
 
To test theoretical models, X-ray data are needed to
determine the physics of the ICM in the radio relic regions. In this
paper we analyze XMM-Newton data of the Coma radio relic 1253+275.
This source is the prototype of cluster relic sources (Giovannini et
al. 1991 and references therein).  It is located at the Coma cluster
periphery in the SW direction, beyond the galaxy group associated with
NGC~4839, at $\sim$ 75\arcmin~ from the cluster center. This location 
corresponds to a projected distance of $\sim$ 2.1 Mpc.  The structure 
of the Coma radio relic is elongated
with the major axis of $\sim$ 840 kpc and is roughly tangential to the Coma
cluster radius.  The integrated spectrum is fairly straight between
150 MHz and 5 GHz, with a spectral index $\alpha$ = 1.18 
(Thierbach et al. 2003).

We performed an imaging and spectral analysis of archival XMM-Newton
data, to derive the properties of this relic and to shed more light
on its physical origin. This is the first detailed study of the X-ray
properties of a radio relic.

We assume the concordance cosmology with H$_0$=70 km s$^{-1}$
Mpc$^{-1}$, $\Omega_m$ = 0.3, and $\Omega_{\Lambda}$ = 0.7.  At the
redshift of the Coma cluster, z = 0.023, the angular scale of
1\arcmin~ corresponds to a linear size of 28 kpc.

\section{XMM-Newton observations and data treatment}

The region of the Coma radio relic was observed with XMM-Newton for a
total of 22~ksec on June 10 2003.  We retrieved the data from the
archive and pipeline processed them using SAS version 6.5. We followed 
the guidelines in the XMM-Newton ABC-guide (Snowden et al. 2004) for the
data preparation.  We screened the XMM-Newton
observation for time intervals of high background emission (flares),
which can contaminate XMM-Newton observations. We examined lightcurves
of the observations in two energy bands: first in the 10-12 keV (12-14
keV) energy band for MOS (pn) data (see also Majerowicz et al. 2004)
and subsequently in the 0.3-10 keV energy band (see Nevalainen,
Markevitch \& Lumb 2005 for details on flares). After the flare
selection, exposure times are of 17(8)~ksec for the MOS (pn) cameras.

We used the weight method for the vignetting correction, attributing a
weight factor to each event (Arnaud et al. 2001).  Since there is
extended low surface brightness emission in the region, we need to
correct for the background. For this we used blank sky observations
compiled by A. Read (Read \& Ponman 2003), which we screened for
flares in the same way as in the Coma relic observations
\footnote{We also tried the background files from Nevalainen, Markevitch
\& Lumb (2005); however, using these files after additional flare screening 
we obtained fit results with  $\chi^2$ values higher than those obtained using 
Read's background files.}.

\section{Results}

\subsection{Imaging analysis}

\begin{figure}
\centerline{
\psfig{figure=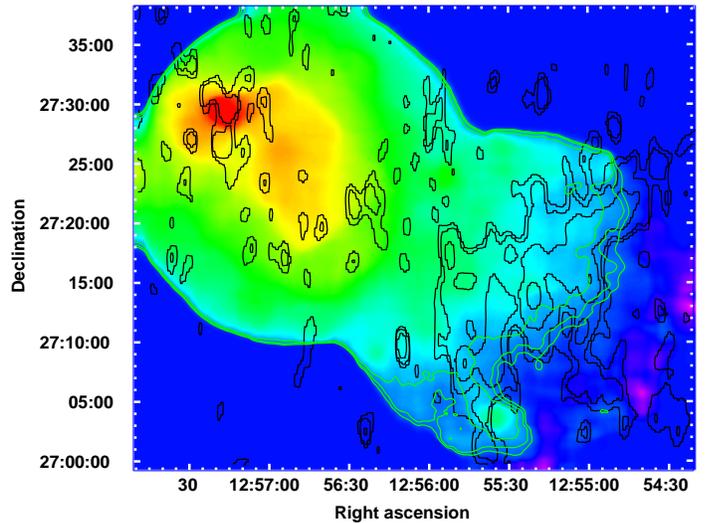,clip=,width=9.5cm}}
\caption
{Color-scale X-ray emission of the NGC~4839 group in the 0.3 $-$ 2 keV
energy range obtained from the MOS cameras.  The image,  
obtained from a mosaic of two pointings, has been smoothed with a 
gaussian filter with $\sigma$ = 10\arcsec, and further with a median filter.  
Dark blue areas are regions with no X-ray data. The green contours represent
the X-ray brightness at 8, 1.6, and  2.4 $\times$ 10$^{-7}$ cts s$^{-1}$
pixel$^{-1}$ (pixel=4.4\arcsec $\times$ 4.4\arcsec), which correspond to
2, 3, and 5 $\sigma$ level over the background.  The radio emission from
the relic at 90 cm is shown by the black contours, whose levels are
2.5, 4, 8, and 16 mJy/beam (beam = 55\arcsec$\times$125\arcsec,
RA$\times$DEC).
}
\label{xmm1}
\end{figure}

We chose for the imaging analysis the energy band 0.3-2 keV, since it
allows the best signal-to-noise ratio (see also Scharf 2002). Figure
\ref{xmm1} shows the filtered count-rate image of the MOS cameras in
this energy band.  Two XMM-Newton observations were merged together
for this image: the data taken in the PV-phase of the sub-group around
NGC~4839 (Neumann et al. 2001) and the
observations described here. The PV-phase data (latest pipeline
processing) were treated in the same way as the Coma relic
observations. Figure \ref{xmm1} is vignetting-corrected and 
background-subtracted using the
above-mentioned blank sky observations. As can be seen
from the image, the X-ray emission around the sub-group extends into the radio
relic region and shows a strikingly similar outer boundary.  It is
clear that the X-ray emission in the relic region comes from the
sub-group. It cannot originate from the Coma cluster itself, as seen when
extrapolating the beta-model derived by Briel, Henry \& B\"ohringer
(1992). In fact, the Coma cluster emission at these radii is more then 20 times
fainter than what is observed.

\begin{figure}
\centerline{
\psfig{figure=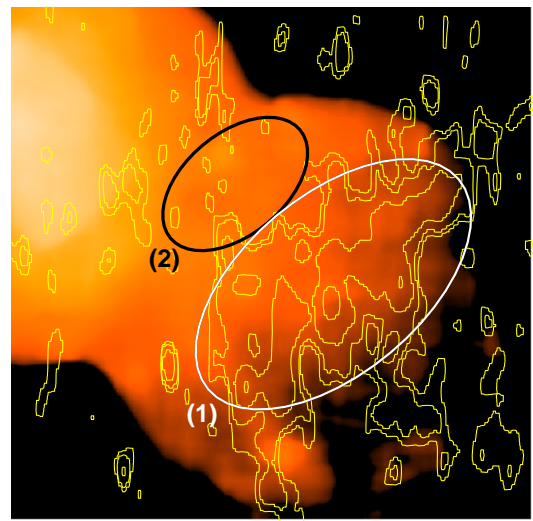,width=7.0cm}}
\caption
{Zoom of Fig.\ref{xmm1} on the region around the radio relic  1253+275. The radio contours are identical to Fig.\ref{xmm1}. The ellipses indicate the regions used for the spectral analysis.
}
\label{xmm2}
\end{figure}

\subsection{Spectral analysis}

To check for the presence of a shock wave and of non-thermal emission, we
selected two regions for the spectral analysis (Fig. \ref{xmm2}):
one ellipse  of 600 $\times$ 330 kpc$^2$ in the relic region (1) and
another ellipse of 310 $\times$ 190 kpc$^2$ to the NE of the relic 
towards the Coma cluster (2).
The total number of source photons in the
0.3-10~keV band in regions (1) and (2) is 13520 and 10620,
respectively.  In these regions, the source count-rate in the
0.3-10~keV energy band is similar to the background count-rate.

Table \ref{tab:spec} shows the results of the spectral fitting
analysis with XSPEC.  For this study, we used the latest response
matrix files avalable at ESA's XMM-Newton homepage and the on-axis
auxiliary reponse files describing the effective area of each camera
as a function of photon energy. We chose the corresponding background
regions in the same detector coordinates. There is an intensity
variability in the instrumental background of XMM-Netwon. To correct
for this effect, we compared the source and background count-rates in
the 10-12 keV (12-14) keV energy band for the MOS (pn) cameras over
each entire camera (see Majerowicz et al. 2004 for details). The ratio
of the count-rates is used as the normalization factor of the
background to the source spectrum. The normalization value in our case
is about 1.2, which indicates that the instrumental background in our
observation is higher than the instrumental background in the blank
sky observations. The chosen background 
normalization of 1.2 provides the lowest $\chi^2$
results.  Varying the normalization factor, the changes of the
spectral fitting results are always within the error bars.

\begin{table}
\caption{Spectral fit results in the relic region (1) and in the
region NE of the relic (2). For the thermal component we use XSPEC's
MEKAL-model with free metallicity and hydrogen column density
(nH). }
\begin{tabular}{cccccc}
\hline
Reg. & model & kT/ $\alpha$ & nH & & red. $\chi^2$\\ & & keV / &
$10^{20}$cm$^{-2}$ & & \\
\hline
\hline
1 & th. & $3.3^{+0.3}_{-0.4}$ & $<0.65$ &  & 1.16 \\
1 & th./ n-th. & $3.2^{+0.5}_{-0.3}$/2.2 & $<0.01$ & & 1.17 \\
2 & th. & $3.3_{-0.5}^{+0.7}$ & $0.58_{-0.58}^{+1.42}$ &  & 1.41 \\
\hline
\end{tabular}

Note. Confidence intervals of the parameters are at 90\% level.
\label{tab:spec}
\end{table}

We fit both a thermal model and a combined
thermal/non-thermal model with a fixed power law index of 2.2, derived
from the radio spectrum (see Sect. 4.2), to the relic region.  
Both fits give consistent
results and point to a thermal origin of the X-ray emission, since the
combined fit gives a non-thermal component contribution consistent
with zero.  A spectral fit with only a non-thermal component of
$\alpha=2.2$ does not provide an acceptable fit, giving a higher
reduced $\chi^2$ and a nH value of $9\times 10^{20}$ cm$^{-2}$, i.e.
an order of magnitude higher than the  observed value of $9\times 10^{19}$
cm$^{-2}$ (Dickey \& Lockman 1990).  Moreover, a thermal-model fit
applied to the region NE of the relic provides a gas temperature fully
consistent with that of the relic region. The relatively high reduced
$\chi^2$ value in this region can be explained by residuals in the
instrumental background subtraction, which are amplified by the
vignetting correction.

The non-thermal emission in the relic region, if at all detectable
with XMM-Newton, would be much lower than the thermal emission. We
calculate the upper limit of this non-thermal component in two ways:
a) spectral fitting, which gives a maximum flux in the 0.3 $-$ 10~keV
band of $2.5\times 10^{-13}$ erg cm$^{-2}$ s$^{-1}$, and b) imaging
analysis. For the latter, since
 no excess X-ray emission directly related to the relic is
detected, we can calculate how much of this emission can be
``hidden''. Assuming Poisson statistics, we estimate that this is
$3.2\times 10^{-13}$ erg cm$^{-2}$ s$^{-1}$, which is in excellent
agreement with the value obtained from the spectral analysis. We adopt
as an upper limit of the non-thermal component
the more conservative value, i.e. $3.2\times 10^{-13}$ erg cm$^{-2}$
s$^{-1}$.

\section{Discussion}

The results of this paper can be summarized as follows:
\par\noindent
a) There is extended X-ray emission in and around the Coma relic
region, which is connected to the sub-group around NGC~4839.
The group is on its first infall onto the cluster, with velocity
$\sim$ 1700 km s$^{-1}$, as suggested from optical data by
Colless \& Dunn (1996), and as further demonstrated by the
effects of ram pressure stripping detected from XMM-Newton data that
indicate the infall direction (Neumann et al. 2001).
\par\noindent
b) The best-fit gas temperatures in the region of the relic and in its
vicinity (NE region) are very similar, around 3.3 keV. This value is
lower than the $4.4\pm0.4$ keV found in the main body of the
NGC~4839 sub-group (Neumann et al. 2001), although the difference is
marginally significant. However, what is most remarkable is that
we do not detect any high temperature gas, which could indicate
the presence of a shock at the location of the Coma relic.
\par\noindent
c) The emission of the relic is largely dominated by thermal
emission. This is deduced by the spectral fit analysis and is also
supported by the following arguments: i) the X-ray emission is
connected to the hot gas of the NGC~4839 sub-group, and ii) there are
no distinct X-ray features connected to the radio relic itself. 
We can only set a flux upper limit of
$3.2\times 10^{-13}$ erg cm$^{-2}$ s$^{-1}$ in the 0.3-10~keV band
for the non-thermal component.

A brief discussion of the implications of our results follows.

\subsection{Merging process and relic formation}

According to current models, the relativistic electrons radiating in
radio relics are expected to be accelerated by shocks originating from
cluster mergers. In a fully ionized plasma, a shock with compression
factor $C$ and Mach number $M$ can accelerate particles to a power law
distribution
in momentum, with slope $\delta = (C+2)/(C-1) = 2(M^2+1)/(M^2-1)$
(Drury 1983).  The spectral index $\alpha$ of the radio emission is
thus related to the Mach number of the shock by the relation: $\alpha
= (M^2 + 3) / 2(M^2 - 1)$.  In this framework, the Coma relic with
spectral index $\alpha$ = 1.18 should originate from particles
accelerated by a shock of Mach number M $\sim$ 2.  
In the case of a monoatomic gas, the ratio
between the post-shock and pre-shock temperatures $T_2$ and $ T_1$ is
given by  $T_2/T_1 = (5M^2 - 1) (M^2 + 3)/16M^2$ (see, for example, 
Sarazin 2002). 
This implies that the temperature jump across a shock of
Mach number M $\sim$ 2 is 2.1, which is not observed in our data.

The above calculations refer to the simplest case.  More rigorously,
one should consider a) the effect of electron ageing on the observed
steep radio spectrum, b) the influence of magnetic field on the shock
parameters, c) the case of re-acceleration of relativistic
particles, leading to possibly higher Mach numbers.

No temperature gradient is detected between the region of the relic
(1) and the pre-shock region (2).  This cannot be due to projection
effects.  In fact the high resolution radio images of the relic
show a sharper brightness decrease on the western (outer) edge
(Giovannini et al.  1991), and the infall path of the
NGC~4839 group to the cluster in a two-body model lies at about
74\degrees~ to the line of sight (Colless \& Dunn 1996). Thus, a
strong inclination of the shock with respect to the sky plane,
which would spread the observed temperature jump over a larger region,
should be ruled out.  The identification of region (2) as the
pre-shock region relies on the assumption that 
the group is on its first passage into the
cluster core. This is a widely accepted scenario (see beginning of
Discussion).  Nevertheless, the low  temperature at the relic location 
is a significant result in itself, in
particular because it is lower than that of the NGC~4839 group, thus
indicating the lack of high temperature shocked gas in the relic
region.

This result favors the hypothesis that turbulence may be the major
mechanism responsible for the physical origin of the radio relic
emission. During its infall onto the Coma cluster, the sub-group 
around NGC~4839 encounters a region of relativistic particles connected
to a magnetic field.  
The interaction between
the ionized moving plasma and the magnetic field would imply
energy transfer from the ICM to the relativistic particles.  As for
the origin of relativistic particles, En{\ss}lin et al. (1998) suggested that
they could be produced by the tailed radio galaxy NGC~4789, located
to the SW of the relic.

The energy for particle acceleration comes from the kinetic
energy of the X-ray gas, which is subsequently slowed down. We see
indications that  the X-ray gas slows down: for example,
the gas of the sub-group
is lagging behind the galaxies of the sub-group (Neumann et al. 2001).
This deceleration is partly due to ram pressure stripping, which acts
more on the extended X-ray gas than on the galaxies.  The total
minimum energy in the radio relic, computed under equipartition
(Thierbach et al. 2003), is $\sim$ 3.5 $\times$ 10$^{58}$
erg, assuming that the radio relic volume can be approximated by an
ellipsoid of 30\arcmin$\times$10\arcmin$\times$10\arcmin,
corresponding to $\sim$ 3.45 $\times$ 10$^{7}$ kpc$^3$. The kinetic
energy of the ICM of the sub-group, assuming a gas mass of $10^{13}$
M$_{\sun}$ and a velocity difference of 1700 km sec$^{-1}$, is $\sim$ $3\times
10^{62}$ erg. Thus, the energy from the group motion can be transferred
to the relativistic electron population with a reasonably low
efficiency.

The head of the NGC~4839 group crossed the relic region $\sim 5
\times 10^8$ yr ago. However, the
relativistic electrons radiating at frequencies between 150 MHz and
1.5 GHz are only  1 $\div$ 2.5 $\times$ 10$^8$ yr old, 
thus supporting the view that
particle reacceleration has occurred more recently.

\subsection{Limit to the magnetic field}

X-ray emission of non-thermal origin is predicted to originate in a
synchrotron source, due to the scattering between the radio-emitting
electrons and the microwave-background photons.  Since synchrotron and IC
emissions are produced by the same electron population, they share
the same emission spectral index $\alpha$.  This spectral index
relates to the index $\delta$ of the power-law electron energy density
distribution as $\delta=2\alpha+1$, and to the photon index $\Gamma_X$
of the IC emission as $\Gamma_X=\alpha + 1$.
Combining the standard formulae of the synchrotron and IC emission
mechanisms (see, for example, the derivation in Govoni \& Feretti 2004), 
a lower limit to the volume-averaged magnetic field can be
obtained from the X-ray flux upper limit for the non-thermal component.

Using a radio flux at 610 MHz of 611 mJy (Giovannini et al. 1991) and
the spectral index of 1.18, we obtain a magnetic field B $>$ 1.05 $\mu
G$.  This value can be compared with the estimates obtained by other
arguments. First, the equipartition magnetic field derived by Thierbach et
al. (2003) by integrating the energy spectrum of particles over all
energies above 300 MeV, and by assuming a filling factor = 1 and a 
proton-to-electron energy ratio k = 1, is 0.67 $\mu$G, scaled to our 
cosmology. Second, the expected value of the magnetic field at the
distance of the relic, derived from the profile of the Coma cluster
magnetic field presented by Brunetti et al. (2001), is of
 about 0.1 $\mu$G.  The profile is obtained in the framework of a model, 
which reproduces the observational properties of the Coma radio
halo invoking electron reacceleration.
Finally, the average magnetic field in the Coma cluster, 
obtained from the
non-thermal hard X-ray emission detected with BeppoSAX is $\sim$ 0.2
$\mu$G (Fusco-Femiano et al. 2004).

The lower limit derived here is almost two times larger than the
equipartition value and is about an order of magnitude larger than the
value derived from the profile presented by Brunetti et al. (2001).
This indicates that the magnetic field in this region of the Coma
cluster has undergone some amplification. This could be related to the
sub-group gas overdensity (Dolag et al. 2001). Another possibility is
that the magnetic field could have been deposited in the ICM by
formerly active radio galaxies, also responsible  for the origin of
relativistic particles.  Since this is a very small region compared to
the whole cluster, a high magnetic field here is not in contrast with
the low average value of the Coma cluster, which should still be
dominated by the magnetic field decline with radius (see also the
discussion in Fusco-Femiano 2004).

\begin{acknowledgements}

We are grateful to R. Fusco-Femiano, G. Giovannini, F. Govoni,
C. Sarazin and M. Tagger for fruitful discussions. We thank the
anonymous referee for useful suggestions.  LF acknowledges a grant
from Egide (Centre fran\c{c}ais pour l'accueil et les \'echanges
internationaux) for a ``S\'ejour scientifique de haut niveau'' at
CEA/Saclay, and thanks the colleagues of the Service d'Astrophysique
for their kind hospitality.

\end{acknowledgements}

\end{document}